\documentclass[sn-mathphys-num]{sn-jnl}

\usepackage{graphicx}%
\usepackage{multirow}%
\usepackage{amsmath,amssymb,amsfonts}%
\usepackage{amsthm}%
\usepackage{mathrsfs}%
\usepackage[title]{appendix}%
\usepackage{xcolor}%
\usepackage{textcomp}%
\usepackage{manyfoot}%
\usepackage{booktabs}%
\usepackage{algorithm}%
\usepackage{algorithmicx}%
\usepackage{algpseudocode}%
\usepackage{listings}%
\usepackage{float}

\theoremstyle{thmstyleone}%
\theoremstyle{thmstyletwo}%

\theoremstyle{thmstylethree}%

\begin{document}

\title[Article Title]{Non-Pyrotechnic Radial Deployment Mechanism for Payloads in Sounding Rockets}


\author[1]{\fnm{Thakur Pranav} \sur{G. Singh}}\email{thakur4pranav@gmail.com}
\equalcont{These authors contributed equally to this work.}

\author[2]{\fnm{Utkarsh} \sur{Anand}}\email{utkarshanand221@gmail.com}
\equalcont{These authors contributed equally to this work.}

\author[1]{\fnm{Tanvi} \sur{Agrawal}}

\author*[3]{\fnm{Srinivas} \sur{G.}}\email{srinivas.g@manipal.edu}

\affil*[1]{\orgdiv{Mechanical \& Industrial Engineering}, \orgname{Manipal Institute of Technology}, \orgaddress{\street{Manipal Academy of Higher Education (MAHE)}, \city{Manipal}, \postcode{576104}, \state{Karnataka}, \country{India}}}

\affil[2]{\orgdiv{Electrical \& Electronics Engineering}, \orgname{Manipal Institute of Technology}, \orgaddress{\street{Manipal Academy of Higher Education (MAHE)}, \city{Manipal}, \postcode{576104}, \state{Karnataka}, \country{India}}}

\affil[3]{\orgdiv{Aeronautical \& Automobile Engineering}, \orgname{Manipal Institute of Technology}, \orgaddress{\street{Manipal Academy of Higher Education (MAHE)}, \city{Manipal}, \postcode{576104}, \state{Karnataka}, \country{India}}}


\abstract{A novel, non-pyrotechnic payload deployment mechanism tailored for sounding rockets is introduced in this research paper. The mechanism addresses the challenge of efficiently and compactly deploying payloads radially during a single launch, featuring a cylindrical carrier structure actuated by a rack-pinion mechanism. Powered by a servo motor, the carrier structure translates to enable radial ejection of payloads. The paper presents the mechanism's design and conducts a comprehensive performance analysis, including structural stability, system dynamics and power requirements. A simulation model is developed to assess payload deployment behavior under various conditions, demonstrating the mechanism's viability and efficiency for deploying multiple payloads within a single sounding rocket launch. The mechanism's adaptability to accommodate diverse payload types, sizes and weights enhances its versatility, while its radial deployment capability allows payloads to be released at different altitudes, offering greater flexibility for scientific experiments. The paper concludes that this innovative payload radial deployment mechanism represents a significant advancement in sounding rocket technology and holds promise for a wide array of applications in both scientific and commercial missions.}




\maketitle

\section{Introduction}\label{intro}

\begin{figure}[!htbp]
\centering\includegraphics[width=0.5\linewidth]{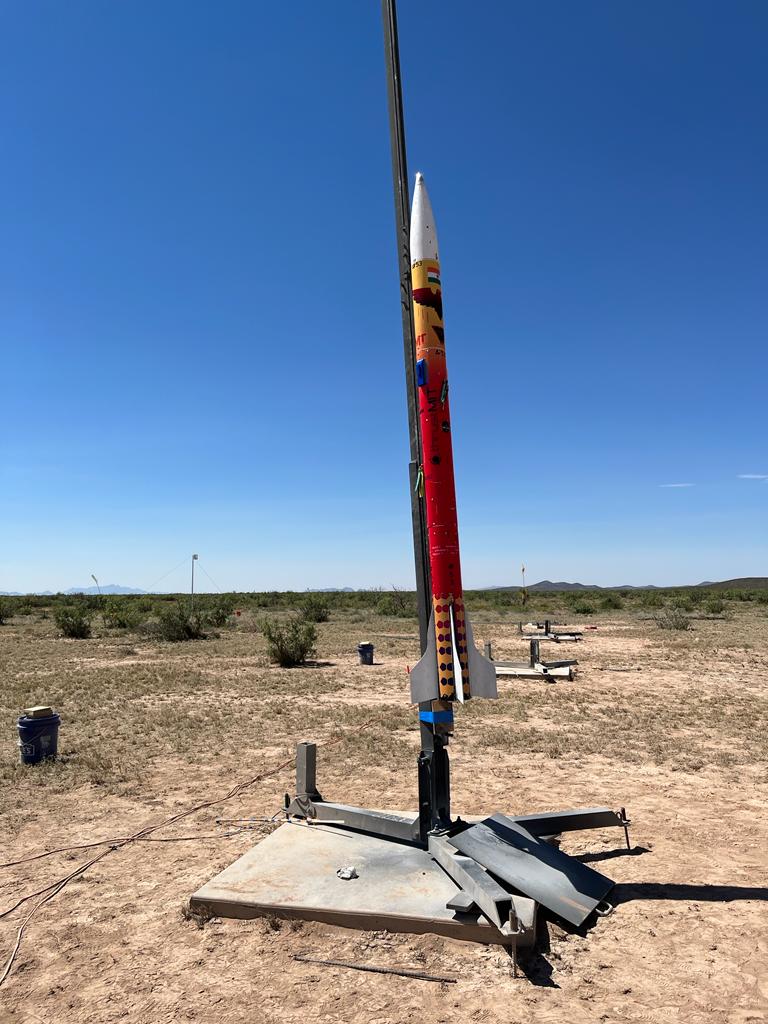}
\caption{thrustMIT's Sounding Rocket: Altair}
\label{altair}
\end{figure}
Sounding rockets are an essential tool for scientific research and exploration, providing a means of collecting data and conducting experiments in the upper atmosphere and beyond \cite{soundingrocket1979}. These rockets are designed to reach high altitudes and provide a brief period of micro-gravity, allowing researchers to study a wide range of phenomena, including atmospheric composition, ionospheric properties, and the behavior of biological and physical systems in a low-gravity environment. Sounding rockets typically consist of a small rocket motor, a payload bay, and a deployment mechanism.[\cite{team65report}] The rocket motor provides the thrust necessary to lift the rocket to the desired altitude, while the payload bay houses the instruments and experiments to be carried out during the flight. The deployment mechanism is used to release the payloads at the appropriate altitude during the rocket flight, allowing researchers to collect data and conduct experiments in real time. One of the key advantages of sounding rockets are their cost-effectiveness and rapid deployment time. Compared to other space launch vehicles, sounding rockets are relatively inexpensive and can be launched quickly and efficiently, allowing researchers to conduct experiments on a relatively tight schedule. Additionally, sounding rockets can be launched from a variety of locations around the world, making them an attractive option for researchers in remote or hard-to-reach areas [\cite{corliss1971nasa}]. thrustMIT's sounding rocket Altair, which was launched in June 2023 at Spaceport America, New Mexico, United States, is shown in \hyperref[altair]{Fig. 1}.

 There are two main types of payload deployment mechanisms: \textbf{Axial} and \textbf{Radial} [\cite{pepermans2022comparison}]. Axial deployment mechanisms are the most common type and are typically used for payloads that require a specific orientation or trajectory. In an axial deployment mechanism, the payload is mounted to a support structure at the top of the rocket and is ejected through a port in the rocket body. The ejection process is typically initiated by a pyrotechnic device, which releases a locking mechanism and allows the payload to be ejected from the rocket. This type of mechanism is ideal for payloads that require a specific orientation, such as telescopes or cameras, and can provide high accuracy and stability during deployment.
 
Radial payload deployment mechanisms represent a significant advancement in the field of sounding rocket technology, offering greater flexibility, versatility, and reliability for scientific experiments. As research continues to push the boundaries of what is possible, it is likely that radial payload deployment mechanisms will become increasingly important and widespread in the field of space exploration and scientific research [\cite{andersson1979survey}].

To establish the contextual framework and underscore the pertinence of this novel mechanism, a comprehensive review of recent literature concerning radial deployment mechanisms will be undertaken. Subsequent sections will delve into the intricacies of the novel mechanism, elucidating its design, operation, and potential for disruptive impact across diverse academic and practical spheres.

\begin{figure}[!htbp]
\centering\includegraphics[width=0.9\linewidth]{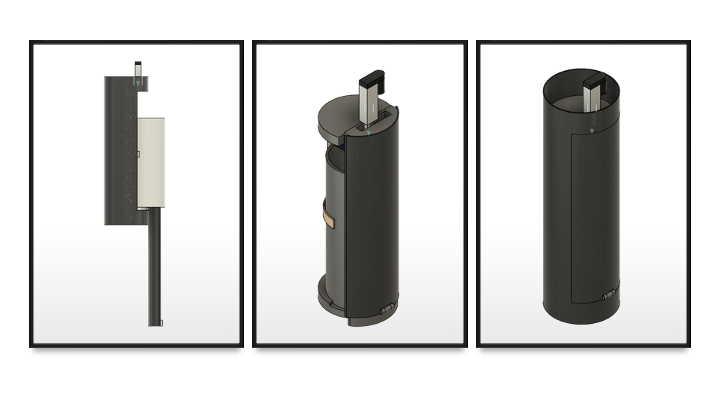}
\caption{Different views of Payload Bay with our Deployment Mechanism}
\label{diff_view}
\end{figure}

\section{Background Theory}\label{bg-theory}
In the realm of engineering and technology, the evolution of deployment mechanisms stands as a testament to innovation in response to the ever-advancing demands of various applications. One such mechanism that has garnered substantial interest is the radial deployment mechanism. Distinguished by its capacity to unfold in a circular pattern, this mechanism offers precision, spatial efficiency, and reliability, characteristics that elevate its significance in a variety of contexts.

Traditional deployment mechanisms, whether linear, rotational, or hybrid, have demonstrated effectiveness in numerous applications. However, they often encounter limitations in scenarios that demand radial mechanism. The extant catalog of radial deployment mechanisms, while serving admirably, faces challenges posed by the evolving complexity of modern technology.

The primary objective of a radial deployment mechanism is to ensure a compact and secure stowage of payloads during the launch and transit phases. Once the spacecraft reaches its designated orbit or location, the mechanism activates, initiating the controlled release of the payload. This deployment process is crucial for achieving proper satellite positioning, optimizing sensor exposure angles, and ensuring successful mission execution.

The theory behind a radial deployment mechanism involves several key considerations:
\begin{itemize}
\item  \textbf{Compact Stowage}: The mechanism must allow the payload to be compactly stored within the spacecraft's confines to optimize space utilization during launch and minimize launch vehicle payload fairing constraints. The design should account for potential volume limitations and weight restrictions [\cite{aiaa20165364}].

\item \textbf{Structural Integrity}: The materials and components used in the radial deployment mechanism should exhibit high strength-to-weight ratios and excellent fatigue resistance. The mechanism must withstand the harsh conditions of launch, space environment, and potential mechanical loads during deployment.

\item \textbf{Reliability and Redundancy}: To ensure mission success and mitigate potential failures, the mechanism often incorporates redundant components and mechanisms. Fail-safe designs and robust deployment sequencing are crucial for the overall reliability of the system.

\item \textbf{Mechanism Actuation}: Radial deployment mechanisms can be actuated through various means, such as pyrotechnics, shape memory alloys, mechanical springs, or motors. The chosen actuation method should be well-suited for the specific mission requirements and should demonstrate repeatability and reliability. 

\item \textbf{Thermal Considerations}: The mechanism must account for thermal variations in the space environment to ensure consistent and predictable deployment performance. Thermal expansion and contraction of materials should be factored into the design to avoid undesirable effects on deployment accuracy.
\end{itemize}

The successful implementation of a radial deployment mechanism plays a pivotal role in a wide range of space missions, including communication satellites, Earth observation satellites, scientific instruments, and interplanetary probes. The efficiency, reliability, and precision of the mechanism contribute significantly to the overall success of the mission and the scientific or operational objectives of the deployed payloads. 

\section{Ongoing Research}\label{research}

The deployment of payloads in sounding rockets has been a critical aspect of scientific research and exploration for decades. Traditional payload deployment mechanisms have primarily relied on axial deployment methods, where the payload is ejected from the rocket's nose cone. While these methods have been effective for many missions, they are not without limitations.

Early sounding rockets employed simple pyrotechnic mechanisms for payload separation. These mechanisms were relatively reliable but lacked precision in controlling the exact moment of deployment. This imprecision could lead to unintended variations in the data collected during experiments. As the demand for more precise scientific measurements increased, so did the need for improved deployment systems. 

In response to these challenges, researchers began to explore alternative deployment mechanisms, including radial deployment systems. The concept of radial deployment involves releasing the payload from the rocket's sides rather than its nose. This approach offers several advantages, including increased payload capacity and improved control over the other process.

Recent developments in radial deployment mechanisms have shown promise. For instance, the use of spring-loaded mechanisms and miniaturized thrusters has allowed for controlled and precise deployment. Researchers have also explored the use of smart materials that can change shape or expand upon activation, providing a novel approach to payload deployment.

Despite these advancements, challenges remain in optimizing radial deployment systems for various payload sizes and mission requirements. The historical evolution of payload deployment mechanisms highlights the need for ongoing research in this field to address these challenges and advance the capabilities of sounding rockets for scientific exploration.

This section provides an overview of the historical context of payload deployment mechanisms and sets the stage for the examination of recent advances and ongoing research efforts in the field.

\section{Problems with other mechanisms}\label{mechanisms}

As mentioned in the previous section, the deployment of payloads can be achieved through two possible means: Axially or Radially. Prior to discussing our radial deployment, it is important to explore the reasons for abandoning axial deployment. Axial Deployment [\cite{doi:10.1109/IEEECONF38699.2020.9389179}], in which the top structure or nosecone is fairing [\cite{MAO2016345}], is the most commonly used method by prestigious space agencies such as ISRO, NASA, and ESA. However, this method presents several technical challenges, particularly with regard to safe pyrotechnic ejection. For orbital or larger sub-orbital rockets, the space constraints are not as significant, thus allowing for numerous pyro-ejecting bolts and systematic wiring. When the nosecones or top halves split into two sections vertically, they are held together by chord wires for sub-orbital rockets, but are left as debris in the case of orbital rockets. This method is relatively expensive due to its constituents of ejection power, precise manufacturing, and small-scale components. Moreover, axial deployment requires redundancy for the ejection mechanism, which is essential but also adds to the costs. For a researcher, the method or mechanism need not be cost-effective, but for a manufacturer or a customer, this is an equally important property. A detailed comparison between various deployment mechanisms is listed in \hyperref[tab:deployment_comparison]{Table 1}.

\begin{table}[!htbp]
\centering
\label{tab:deployment_comparison}
\caption{Comparison of Sounding Rocket Payload Deployment Methods}
\begin{tabular}{|c|p{0.3\linewidth}|p{0.3\linewidth}|}
\hline
\textbf{Deployment Method} & \textbf{Advantages} & \textbf{Limitations} \\
\hline
Axial Deployment &
\begin{itemize}
    \item Simple mechanism
    \item Minimal interference with payload 
    \item Highly accurate and stable deployment of payload.
\end{itemize}
& 
\begin{itemize}
    \item Requires rocket to be in specific orientation
    \item Pyrotechnic charge required depending on payload size.
    \item Multiple payloads cannot be deployed in a single launch.
\end{itemize} \\
\hline
Fairing Deployment & 
\begin{itemize}
    \item Enhanced payload protection during ascent
    \item Suitable for larger payloads 
    \item No specific orientation of rocket required.
\end{itemize}
& 
\begin{itemize}
    \item Increased complexity
    \item Additional weight
    \item Potential aerodynamic constraints 
    \item Explosive bolts used in fairing.
\end{itemize} \\
\hline
Radial Deployment (Favorable) & 
\begin{itemize}
    \item Relatively simpler mechanism
    \item Reduced aerodynamic constraints 
    \item No pyrotechnics involved 
    \item Multiple payloads can be deployed in a single launch 
    \item No specific orientation of rocket required 
    \item Payload protection during ascent.
\end{itemize}
&
\begin{itemize}
    \item Potential risk of tangling.
    \item Weight constraint of usable payloads. Cannot push heavy payloads.
\end{itemize} \\
\hline
\end{tabular}
\end{table}

\section{Deployment Mechanism Design}
\subsection{Mechanical Systems}
Our deployment mechanism uses a single gear Rack and Pinion system to convert rotational motion into linear motion and push the payload radially out. The pinion is connected to a high torque servo motor which rotates at the rated speed. The torque generated by the servo motor is transferred to the connected pinion and is converted into tangential force which acts on the payload through the rack arm.
\begin{figure}[H]
\centering\includegraphics[width=0.45\linewidth]{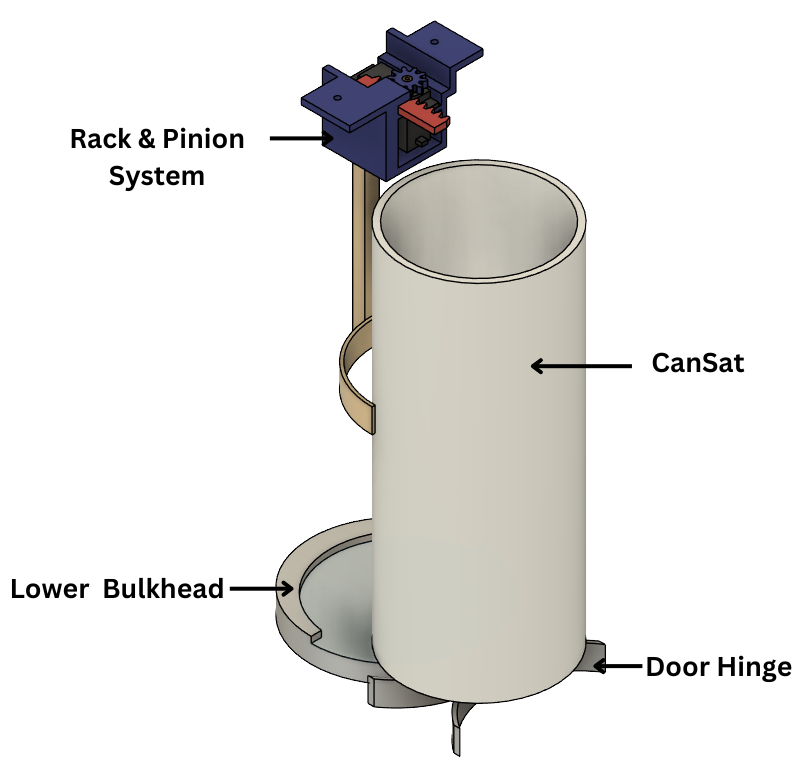}
\caption{Orthogonal view of Deployment Mechanism}
\label{cansat1}
\end{figure}

\begin{figure}[H]
\centering\includegraphics[width=0.45\linewidth]{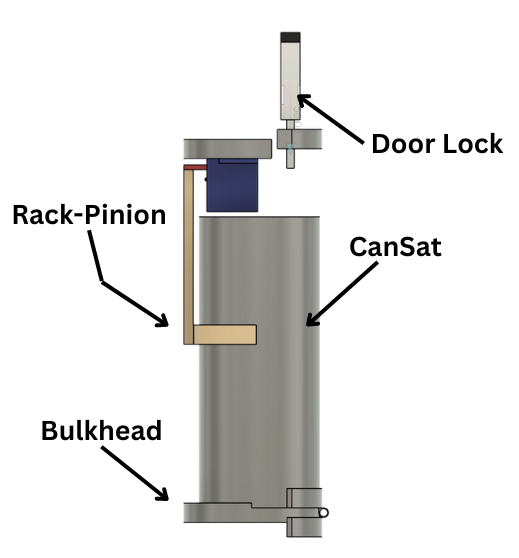}
\caption{Side view of Deployment Mechanism}
\label{cansat2}
\end{figure}

\subsubsection{Rack - Pinion Mechanism}
\begin{figure}[H]
\centering\includegraphics[width=0.8\linewidth]{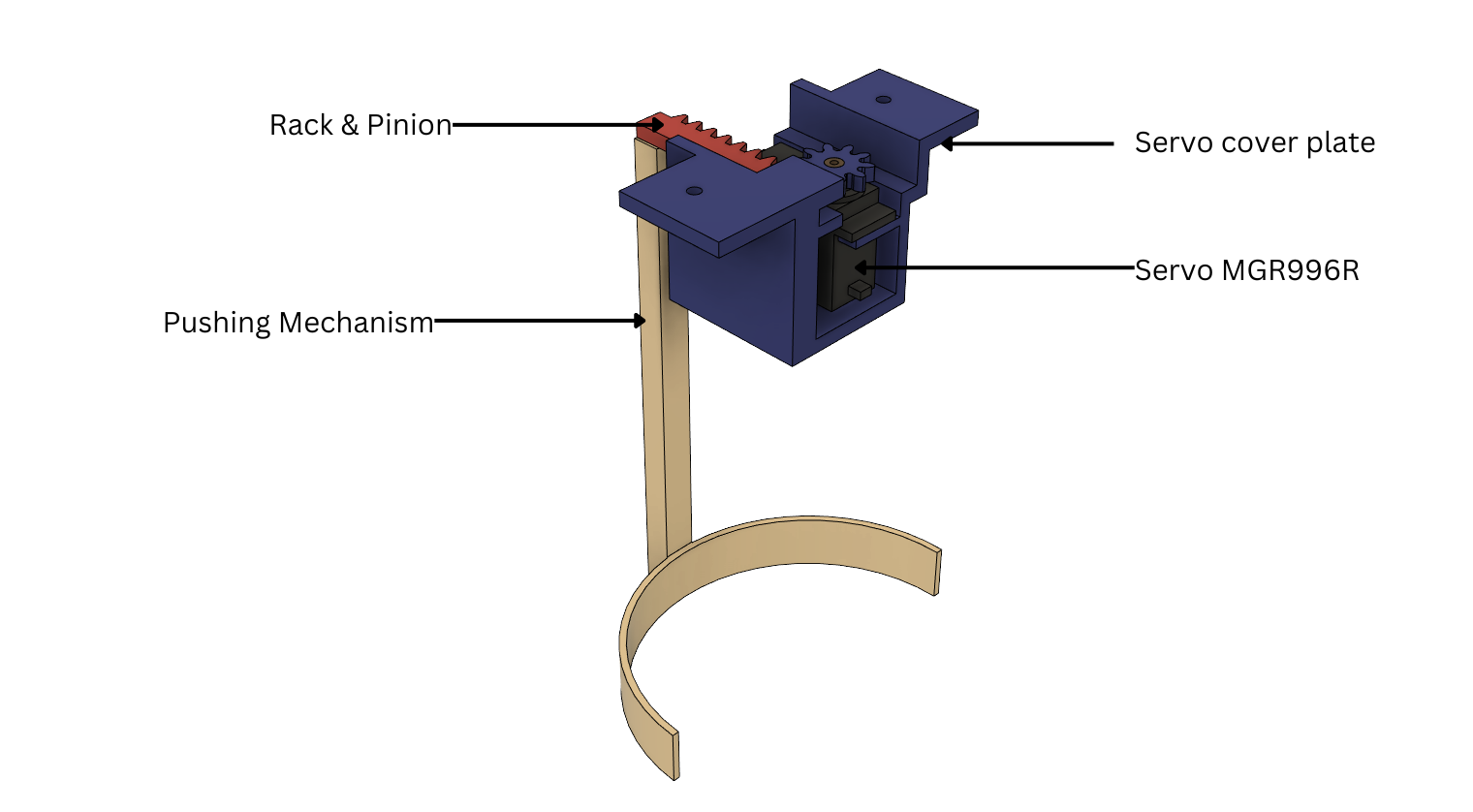}
\caption{Orthogonal view of Rack and Pinion Mechanism}
\label{cad2}
\end{figure}

\begin{figure}[H]
\centering\includegraphics[width=0.8\linewidth]{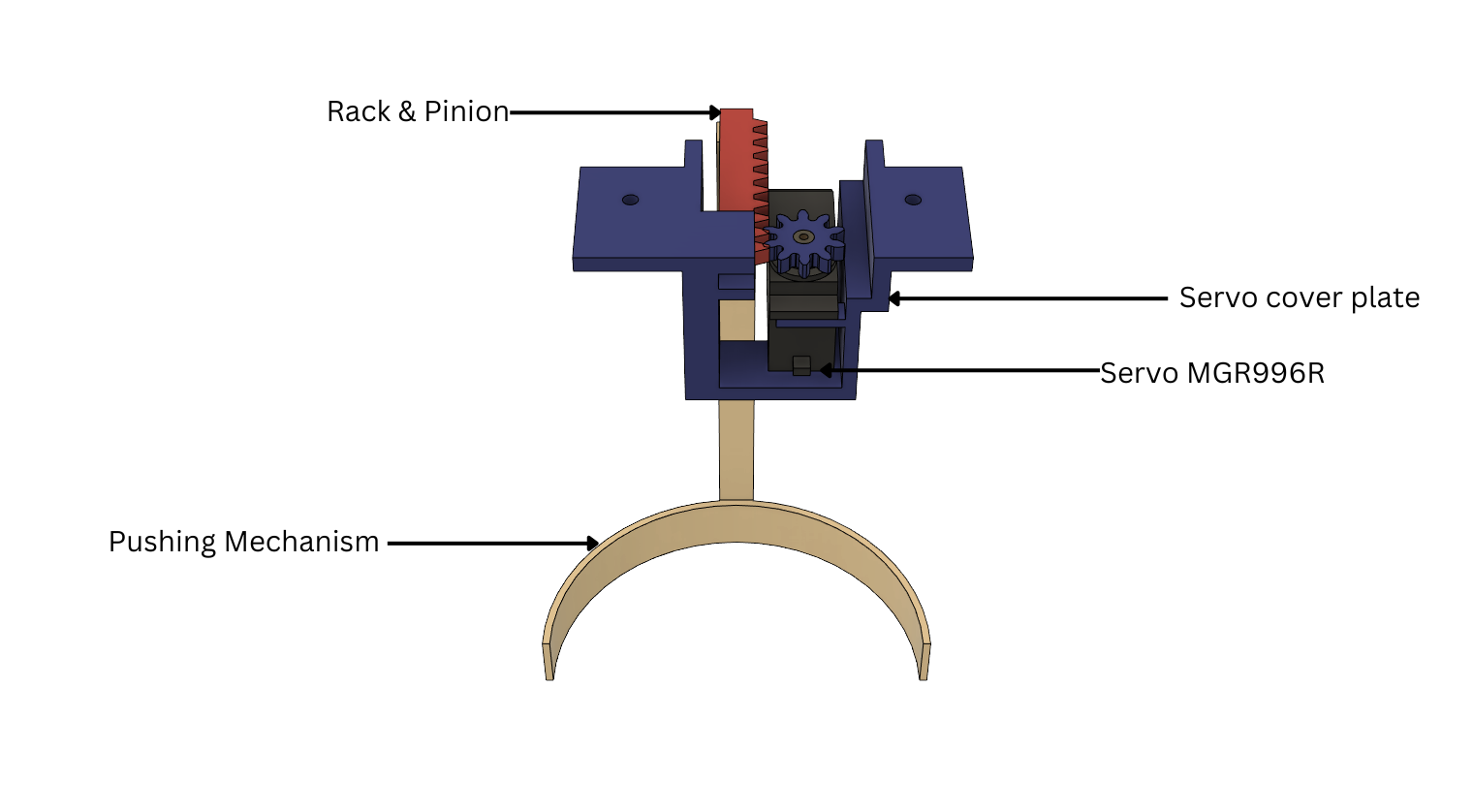}
\caption{Front view of Rack and Pinion Mechanism}
\label{cad1}
\end{figure}

An Aluminium Pinion is connected to a high torque servo motor which sits on a 3D printed servo plate bolted to the upper bulkhead of the payload bay made from ABS plastic. Aluminium is used due to its lightweight and high structural strength properties making them easy to manufacture at the same time. The servo motor rotates at its rated speed, generating a torque which is transferred to the pinion. This torque converted is into a tangential force on the rack through the pinion - rack interface. A semi-circular arm is connected to the rack and when the tangential force acts on the rack, it moves in the horizontal direction moving the arm as well applying force on the CanSat [\cite{chun2023cansat}] pushing it out radially.

\begin{table}[!htbp]
\caption{Gear Dimensions}
\centering
\begin{tabular}{|l|c|} 
 \hline
 Gear Module & 2\\
 \hline
 Pitch Diameter & 20mm (0.79 inch)\\
 \hline
 Root Fillet Radius & 1.1mm (0.04 inch)\\
 \hline
 Gear Thickness & 5mm (0.2 inch)\\
 \hline
 Hole Diameter & 6mm (0.24 inch)\\
 \hline
 No. of Teeth & 10\\
 \hline
 Gear Thickness & 5mm (0.2 inch)\\
 \hline
\end{tabular}
\label{gear_dim}
\end{table}

\begin{table}[!htbp]
\caption{Rack Dimensions}
\centering
\begin{tabular}{|l|c|} 
\hline
Diametrical Pitch & 10 Dp - 14.5°\\
 \hline
 Length & 66.48mm (2.62 inch)\\
 \hline
 Mating Section Height & 4.274mm (0.17 inch)\\
 \hline
 Addendum & 2.137mm (0.08 inch)\\
 \hline
 Dendum & 2.137mm (0.08 inch)\\
 \hline
\end{tabular}
\label{rack_dim}
\end{table}

\subsubsection{Payload Bay Bulkhead}
\begin{figure}[H]
\centering
\includegraphics[width=0.45\linewidth]{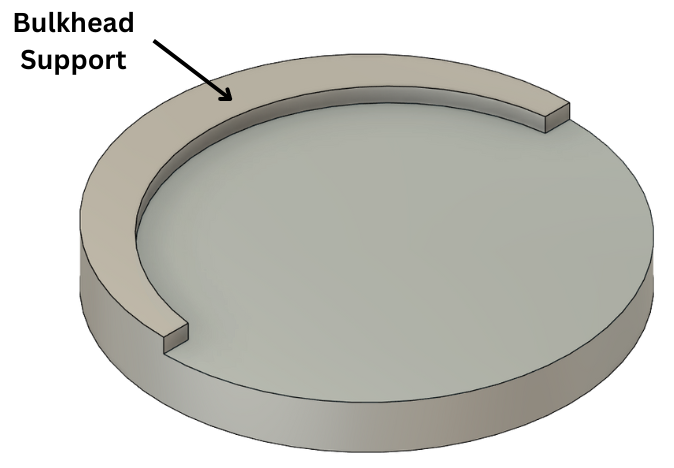}
\caption{Lower Payload Bay Bulkhead}
\label{lower_bulkhead}
\end{figure}
Bulkheads are used at the top and bottom end of the payload bay to maintain the structural integrity of the rocket body helping in efficient load transfer in the rocket withstanding the immense force acting on the rocket during launch , flight etc. They also separate the payload section from the rest of the rocket, protecting delicate instruments and experiments from the harsh environment of launch. The bulkheads are made from Aluminium metal as they are lightweight, have high structural strength and are easy to manufacture.

The payload sits on the lower bulkhead surrounded by a support structure which prevents the lateral displacement of the CanSat during the launch, flight and recovery of the rocket. 

\subsubsection{Payload Bay Door}
\begin{figure}[H]
\centering
\includegraphics[width=0.5\linewidth]{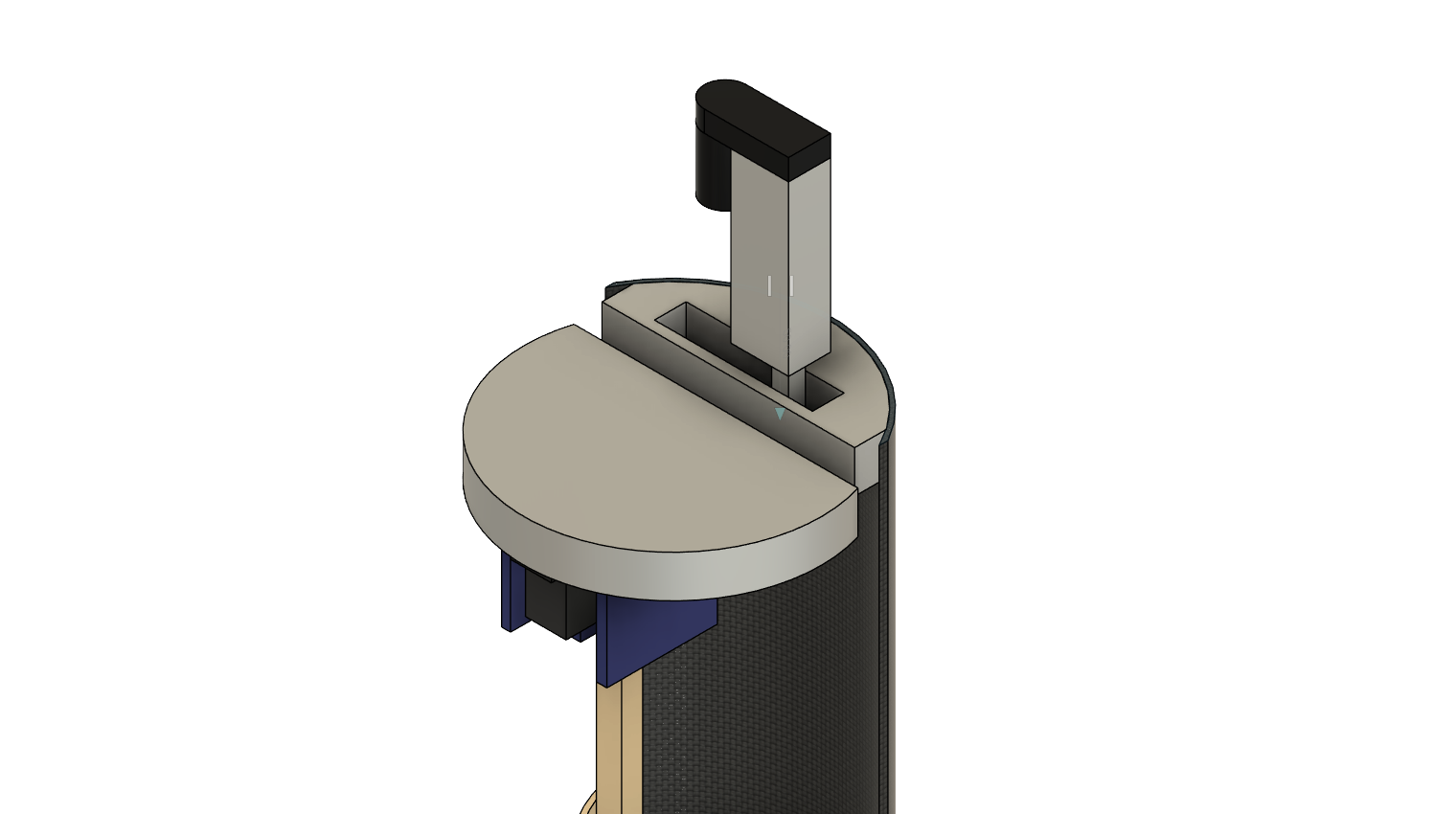}
\caption{Door Lock Mechanism}
\label{pay_lock}
\end{figure}

\begin{figure}[H]
\centering
\includegraphics[width=0.3\linewidth]{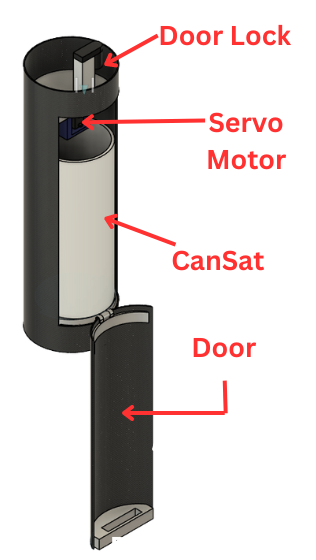}
\caption{Payload Bay with Door Open}
\label{pay_door}
\end{figure}
The door mechanism is implemented on the outer body of the payload bay which prevent the payload from falling out before its scheduled deployment. The door is connected to the rocket body through a hinge at the bottom of the door. A linear actuator which is controlled by a microcontroller is used to lock the door preventing it from opening during rocket flight. 

\subsection{Electrical and Control Systems}
\begin{figure}[H]
\label{elec_block}
\centering
\includegraphics[width=0.5\linewidth]{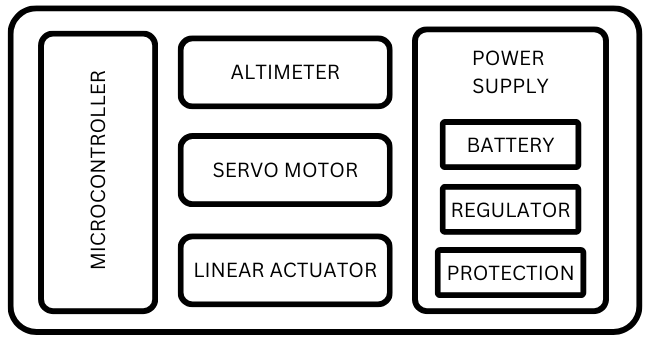}
\caption{Block Diagram of Electrical System}
\end{figure}

\begin{figure}[H]
\label{rtos}
\centering
\includegraphics[width=0.4\linewidth]{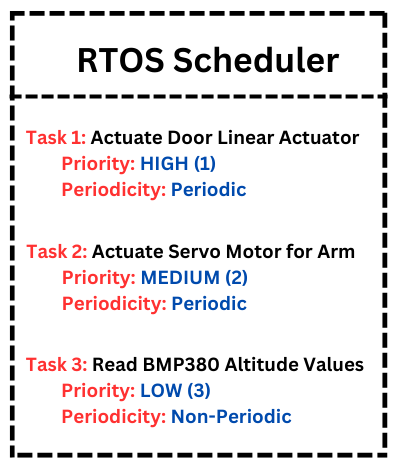}
\caption{RTOS Scheduler}
\end{figure}

The Electrical System is designed for reliable and effective control of the deployment mechanism. It is responsible for opening the payload door and controlling the servo motor to push the payload out when the desired altitude is reached by the rocket. Subsequently, the system also monitors the various flight parameters of the rocket such as pressure, differential pressure, altitude, speed etc. A microcontroller is used as the main processing and controlling unit of the system on which a Real Time Operating System (RTOS) is implemented to optimize its functioning. See \hyperref[elec_block]{Fig. 10} and \hyperref[rtos]{Fig. 11}.

\subsubsection{Processing Unit}
\begin{figure}[H]
\centering
\includegraphics[width=0.4\linewidth]{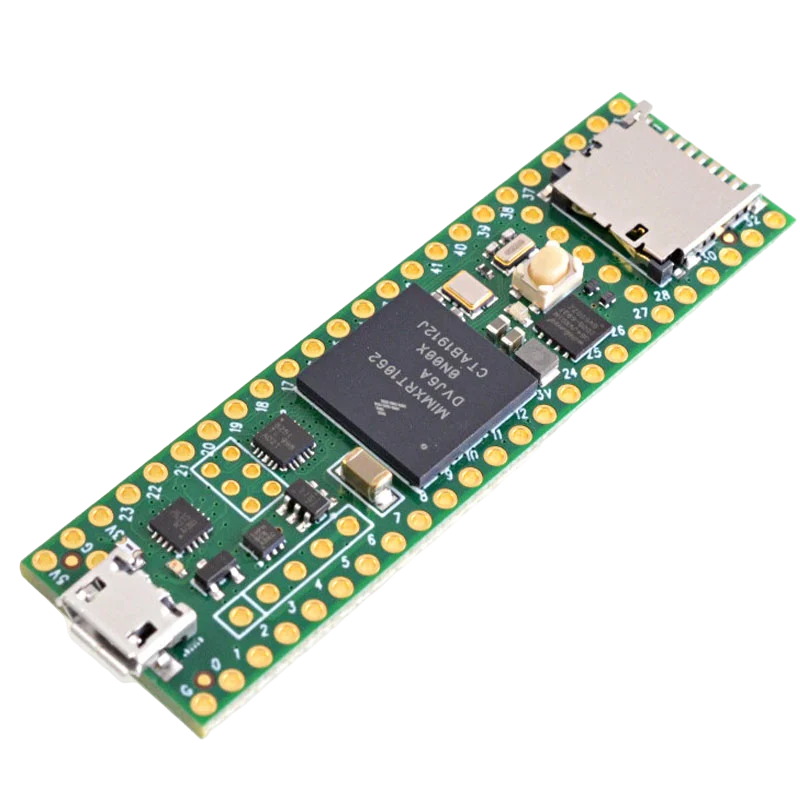}
\caption{Teensy 4.1 MCU}
\label{teensy}
\end{figure}
The system uses a Teensy 4.1 Microcontroller as the central processing unit of the electrical and control system. Teensy 4.1 is a industry standard microcontroller featuring an ARM Cortex-M7 processor which excels at high speed processing of data and is a popular choice for control and monitor applications.

\subsubsection{Measuring Flight Parameters}
\begin{figure}[H]
\centering
\includegraphics[width=0.4\linewidth]{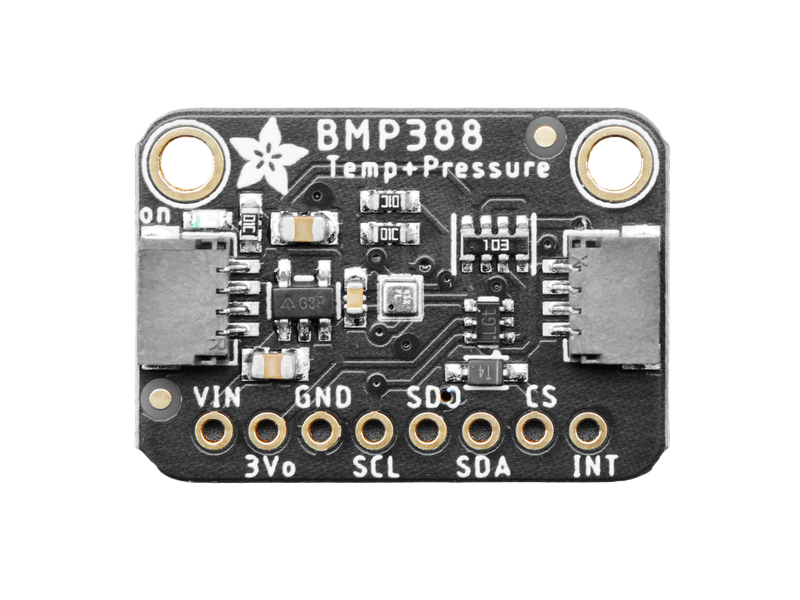}
\caption{BMP380 Altimeter}
\label{bmp}
\end{figure}
The system uses a BMP380 barometric pressure sensor to measure the atmospheric pressure and altitude of the rocket during its flight. BMP380 is a highly accurate, low-cost, low-power and compact design MEMS sensor which can be easily interfaced and integrated into electronic circuits and can be operated into a wide temperature range.

\subsubsection{Actuators}
\begin{figure}[H]
\centering
\includegraphics[width=0.4\linewidth]{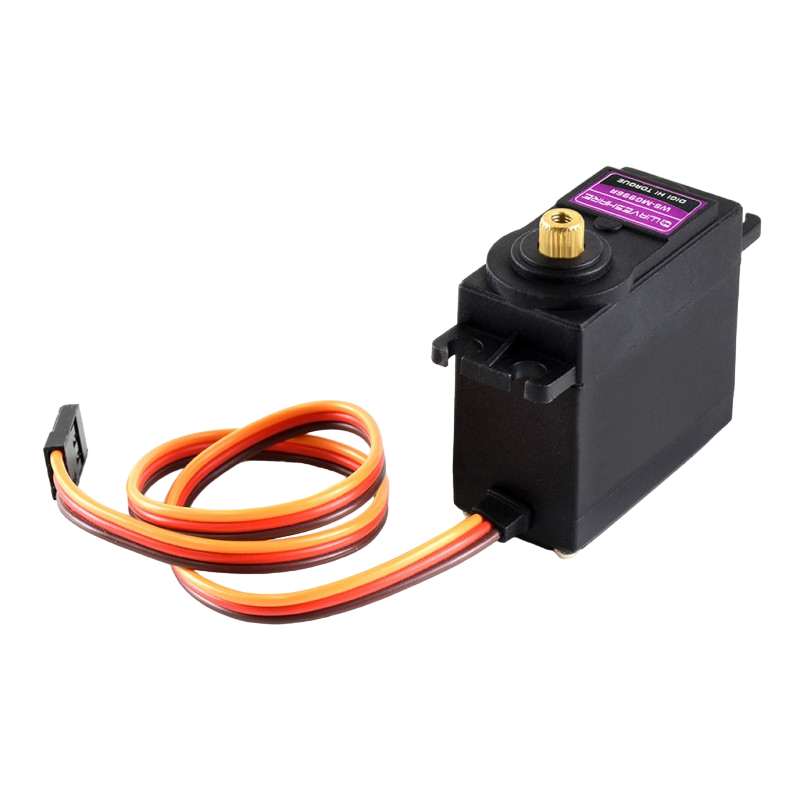}
\caption{MG996R Servo Motor}
\label{mgr}
\end{figure}
\begin{enumerate}
    \item[] \textit{\underline{Servo Motor:-}}
    \item[] We are using a MGR996R high torque digital servo motor to control the pushing arm. It can rotate a full 360° in either directions and features a metal gear resulting in high stalling torque, a double ball design increased durability and an upgraded IC control system for accurate and precise rotations. 
    \begin{figure}[H]
\centering
\includegraphics[width=0.5\linewidth]{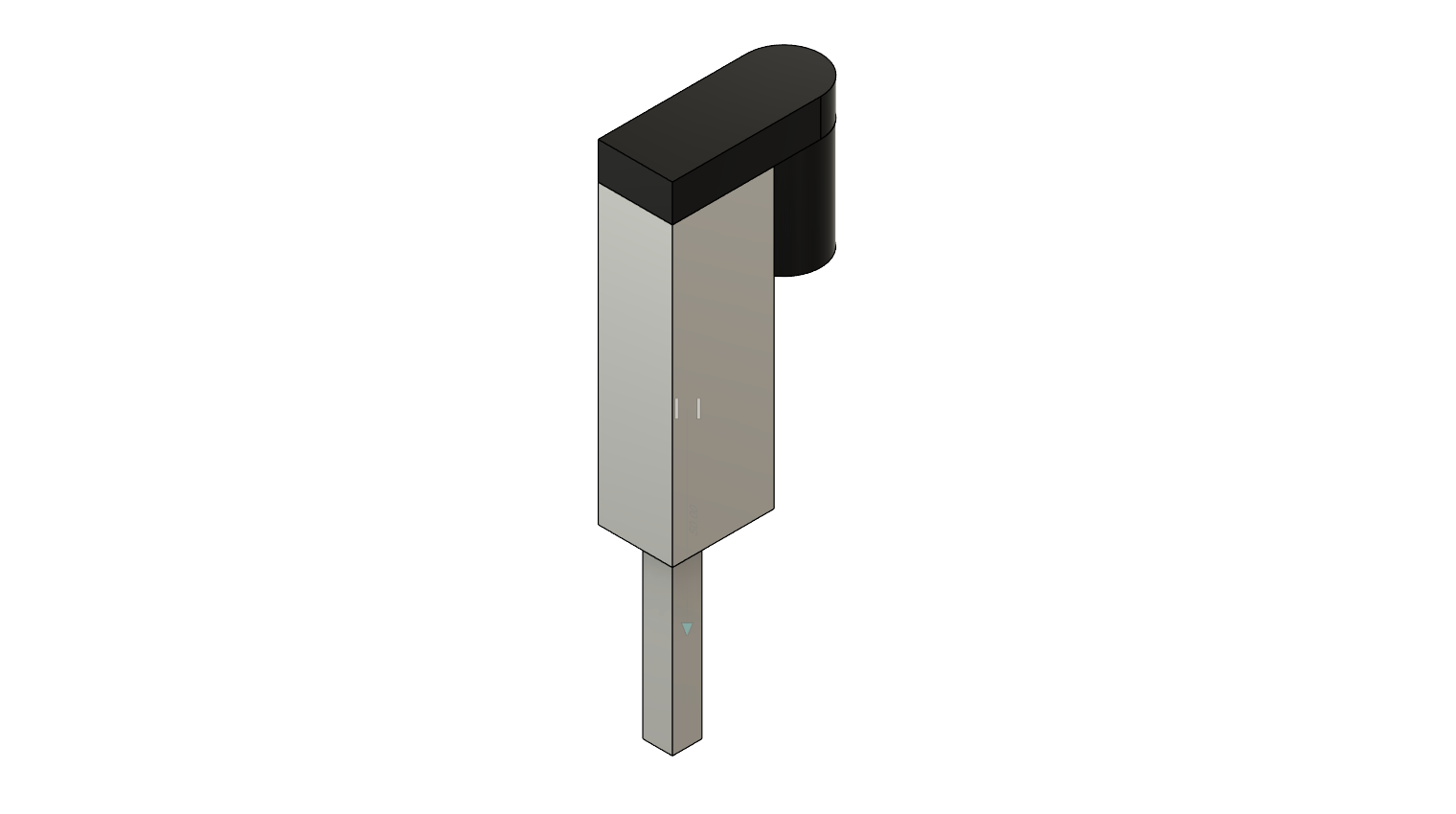}
\caption{Door Linear Actuator}
\label{door_actuator}
\end{figure}
    \vspace{5pt}
    \item[] \textit{\underline{Linear Actuator:-}} 
    \item[]  The system uses a Electric Push Rod Linear Actuator for locking the payload door controlled by the microcontroller. When the door is locked, the push rod of the linear actuator is inserted into a rectangular hole which is made into the top bulkhead. The push rod retracts back when the door is unlocked allowing it to freely move and opening the door due gravity and drag force on the rocket.
\end{enumerate}

\subsubsection{Power System}
\begin{figure}[H]
\centering
\includegraphics[width=0.5\linewidth]{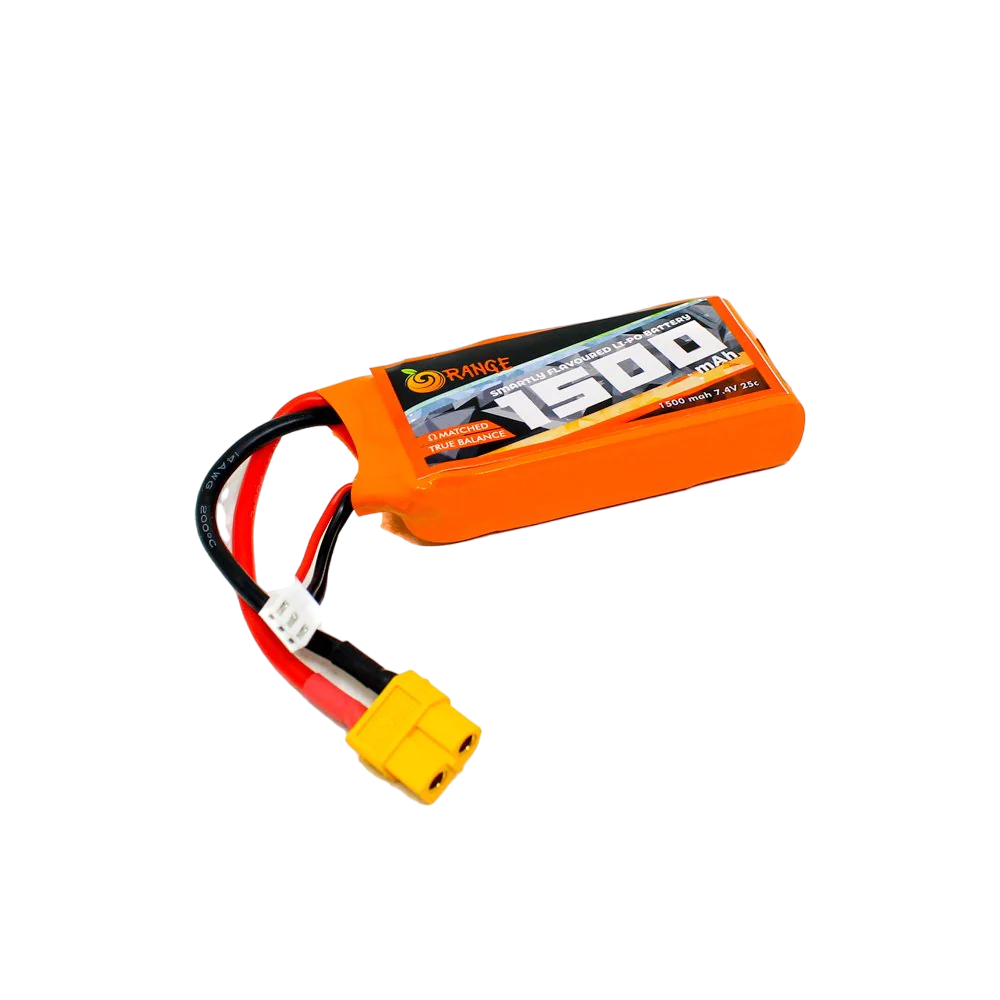}
\caption{2S-7.4V LiPo Battery}
\label{lipo}
\end{figure}
The deployment mechanism is powered by a 2S-7.4V Lithium Polymer (LiPo) battery pack. Voltage generated by the battery is stepped down to desired operating voltages of various electronic components of the system using multiple voltage regulator circuits.

\begin{table}[!htbp]
\caption{Servo Motor Specifications}
\label{motor_spec}
\centering
\begin{tabular}{|l|c|} 
\hline
Weight & 55g\\
\hline
Operating Voltage & 4.8V - 7.2V \\
\hline
Running Current & 500mA - 900mA\\
\hline
Stall Torque(6V) & 10kgf.cm\\
\hline
Stall Current(6V) & 2Amps\\
\hline
Average Operating Speed (6V) & 60 RPM\\
\hline
\end{tabular}
\end{table}

\section{Calculations for Mechanism}

\begin{figure}[H]

\centering
\includegraphics[width=0.6\linewidth]{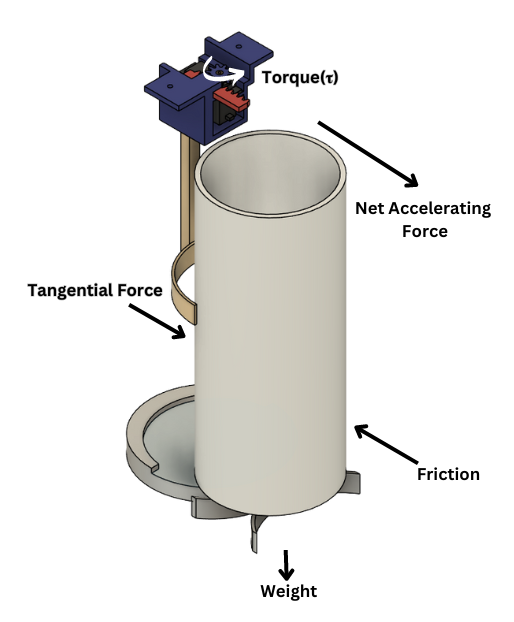}
\caption{Forces Acting on Payload}
\label{forces}
\end{figure}

\subsection{Torque of Servo Motor}
According to the datasheet of the servo motor, the maximum Torque ($\tau$) generated by the servo motor at 6V is $10 kgf.cm$ or $0.98 N$, refer \hyperref[motor_spec]{Table 4}. 

\subsection{Tangential Force on Payload}
\label{tan_force}
The Tangential Force ($F_{t}$) generated on the rack arm which is applied on the CanSat can be calculated using the following formula:-

\begin{equation}
    F_{t} = \frac{2 * \tau}{d}
\end{equation}

Where,
\begin{itemize}
    \item[]  $\tau$: Torque generated by the servo motor which is transferred to the pinion. Refer \hyperref[motor_spec]{Table 4}.
    \item[] $d$: Pitch Diameter of the Pinion. Refer \hyperref[gear_dim]{Table 2}.
\end{itemize}

The maximum Tangential Force is calculated as $98N$. In practical applications, entire energy or power is not $100\%$ transferred between any mechanism and there are various types of mechanical and electrical losses. Considering an industry level efficiency of $85\%$, the actual generated Tangential Force is calculated as $\approx 84N$.

\subsection{Time for Payload Deployment}
\label{time_dep}
An efficient deployment mechanism will eject the payload accurately at the desired deployment altitude. This means that the time from sending the deployment signal, to the actuating of the mechanism to falling of the payload, has to be minimized because the altitude of the rocket is also simultaneously decreasing during the recovery of the rocket and we don't want to miss the desired altitude range. For calculations, an estimated time of $t = 5 s$ is taken by the mechanism to deploy the payload.

\subsection{Horizontal Acceleration of the Payload}
\label{sec_acc}
Due to the tangential force acting on the payload, the payload is accelerated towards the opening on the rocket body covering a distance of $60 mm$ in the horizontal direction before it can fall out. The acceleration ($\alpha$) can be calculated using the following formula:-
\begin{equation}
    S = u*t + \frac{1}{2} * \alpha * t^{2}
\label{eq_acc}
\end{equation}

Transforming the \hyperref[eq_acc]{Eq. 2},

\begin{equation}
    \alpha = \frac{2 * (s - u*t)}{t^{2}}
\end{equation}

Where,
\begin{itemize}
    \item[] $S$: Horizontal distance which the payload has to cover.
    \item[] $u$: Initial horizontal velocity of payload.
    \item[] $t$: Time taken by payload to deploy. Refer \hyperref[time_dep]{Section 6.3}.
\end{itemize}

The acceleration is calculated as $\alpha = 4.8$x$10^{-3} m/s^{2}$.

\subsection{Friction Force on Payload}
\label{friction}
A Friction Force ($f$) acts on the CanSat in the horizontal direction opposite to the direction displacement of the CanSat and Tangential Force. This friction force depends on the mass of the payload and properties of surface which is represented by the coefficient of friction. 

An approximate value of Coefficient of Friction ($\mu$) for aluminum is taken as $0.61$ for the calculations. 

Friction Force can be calculated using the following equation:
\begin{equation}
    f = \mu * m * g
\end{equation}
Where,
\begin{itemize}
    \item[] $\mu$: Coefficient of Friction
    \item[] $m$: Mass of Payload
    \item[] $g$: Acceleration due to gravity
\end{itemize}

\subsection{Maximum Permissible Mass of Payload}
The resultant force (see \hyperref[forces]{Fig. 17}) acting on the payload in the horizontal direction can be represented by the following equation:
\begin{equation}
    F_{t} = f + m * \alpha
\end{equation}
\begin{equation}
    F_{t} = (\mu * m * g) + m * \alpha
\end{equation}

Transforming the equation, the Maximum Permissible Mass of the Payload which can be ejected radially by our deployment mechanism cal be calculated by:
\begin{equation}
    m = \frac{F_{t}}{(\mu * g) + \alpha}
\end{equation}
Where,
\begin{itemize}
    \item[] $F_{t}$: Rack Tangential Force on Payload. Refer \hyperref[tan_force]{Section 6.2}
    \item[] $f$: Frictional Force on Payload. Refer \hyperref[friction]{Section 6.5}
    \item[] $m$: Mass of Payload
    \item[] $\alpha$: Acceleration generated in Payload. Refer \hyperref[sec_acc]{Section 6.4}
    \item[] $\mu$: Coefficient of Friction
    \item[] $g$: Acceleration due to gravity
\end{itemize}

Therefore, the maximum permissible payload mass that can be radially ejected by our deployment mechanism is $\approx \textbf{14 Kg}$.

\section{Testing of Mechanism}
A physical prototype of the mechanism was built using 3D printed parts, metal bulkheads and carbon fiber body tube of the rocket. Multiple deployment tests were conducted on a CanSat made of Glass Fiber weight 1 Kg, to evaluate the performance of the mechanism and its efficiency [\cite{shukla2022satellite}]. No-load tests were done to check whether the tangential force produced by the servo motor was sufficient to push the CanSat out without any hindrance. The force was sufficient for both load (Filled) and no-load (Hollow) condition of the CanSat deploying in the horizontal direction which is perpendicular to the rocket's frame of traversal.[\cite{IAC55th}].

During the initial tests, a few minor issues were identified and analyzed; such as, the gear had interference while rotating with the plate, as the plate compressed upon tightening with the screws by at least 3mm when attached to the top bulkhead. Additionally, the movement between the rack and the rack slot was not as free as it should be, because the 3D printed materials at extrusions needed to be sanded for improved performance [\cite{bulut2013model}].

Further tests involved updates to the previous designs, which included a re-designing of the plate that held the servo and push mechanism to allow sufficient space for gear rotation and rack sliding. The mating of rack and pinion was improved to achieve accurate mating, and lubricants were used to ensure proper sliding of the rack. 

\begin{figure}[!htbp]
\centering\includegraphics[width=0.6\linewidth]{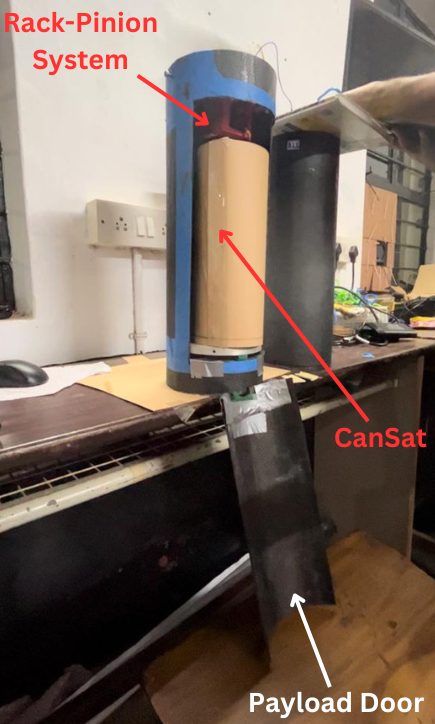}
\caption{Physical Prototype of the Mechanism}
\label{prototype}
\end{figure}

\section{Mechanical Simulations}
The prevailing forces and conditions exhibited by the mechanism obviated the necessity for comprehensive simulations. Hence, structural simulations were performed on the critical load bearing components of the mechanisms. The CAD for the mechanism was modeled in Autodesk Fusion 360 and then imported into Ansys simulation software. The respective materials were configured and the static structural simulation for the model was simulated.

The suitable factor of safety of $1.5$ was considered after the simulations were performed.

\subsection{Structural Simulation on Payload Door }
The payload door hinge component is exposed to atmospheric conditions, exhibiting a minimal extrusion of no more than $3mm$. Subsequently, the payload door will hit the body tube after opening which could cause minor total deformations which were analysed in the simulation.

 The mesh used for the model was hex dominant so to receive the results precisely. The necessary parameters for fixed friction supports were given and the drag force was applied to the exposed regions. 

\begin{figure}[!htbp]
\centering\includegraphics[width=1.00\linewidth]{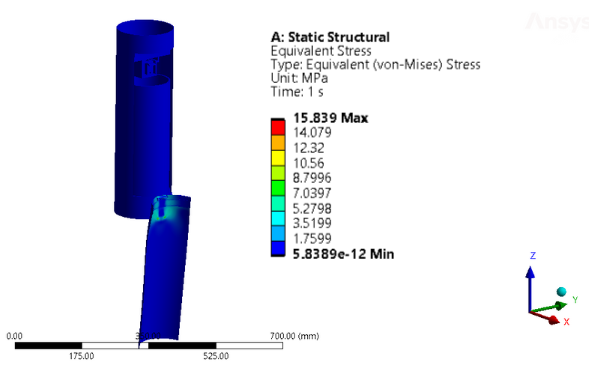}
\caption{Static structural simulation on door representing equivalent Von-Mises stress}
\label{sim1}
\end{figure}

\begin{figure}[!htbp]
\centering\includegraphics[width=1.00\linewidth]{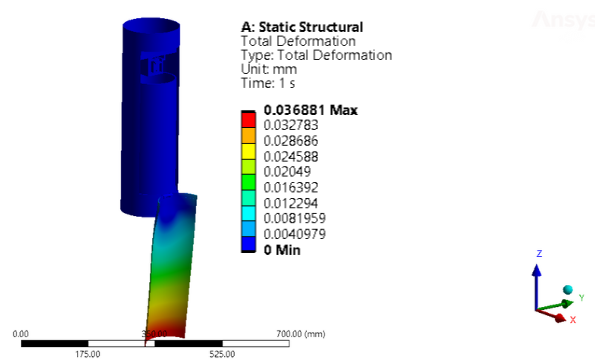}
\caption{Static structural simulation on door representing Total Deformation}
\label{sim2}
\end{figure}

\begin{figure}[!htbp]
\centering\includegraphics[width=1.00\linewidth]{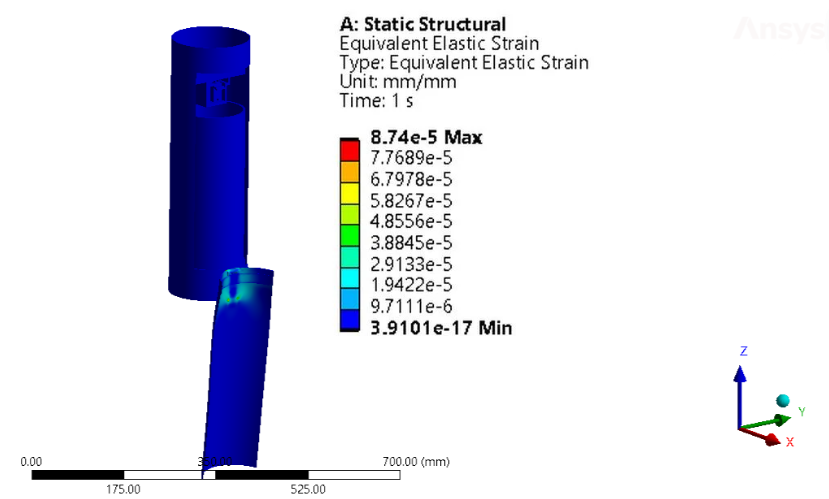}
\caption{Static structural simulation on door representing equivalent Elastic strain}
\label{sim3}
\end{figure}

The application of drag forces to both the hinge and the exposed door facilitated a comprehensive assessment, yielding enlightening results (see \hyperref[sim1]{Fig. 19}, \hyperref[sim2]{Fig. 20} and \hyperref[sim3]{Fig. 21}). The total deformation observed was $0.036 mm$ at the bottom section of the door gradually decreasing whereas the equivalent von-mises stress was observed at the hinge which was approximately $15.839 MPa$. 

\subsection{Structural Simulation on Rack Mechanism}
\begin{figure}[!htbp]
\centering\includegraphics[width=1.00\linewidth]{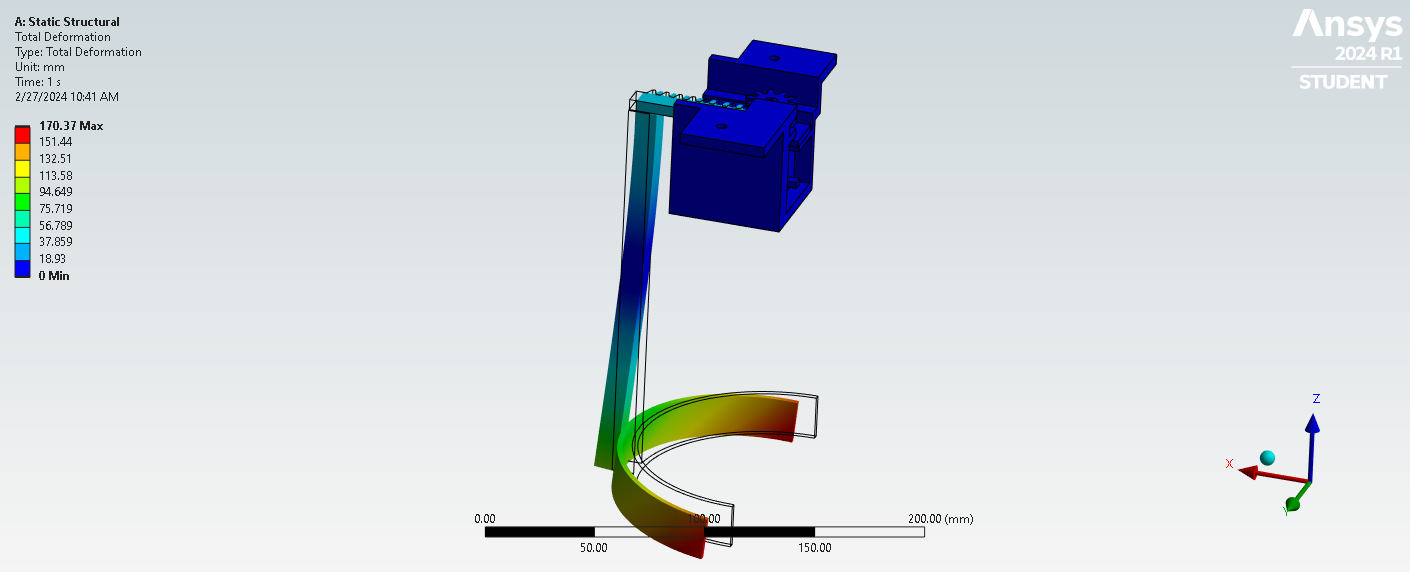}
\caption{Static structural simulation on rack representing equivalent Total Deformation}
\label{racksim1}
\end{figure}

\begin{figure}[!htbp]
\centering\includegraphics[width=1.00\linewidth]{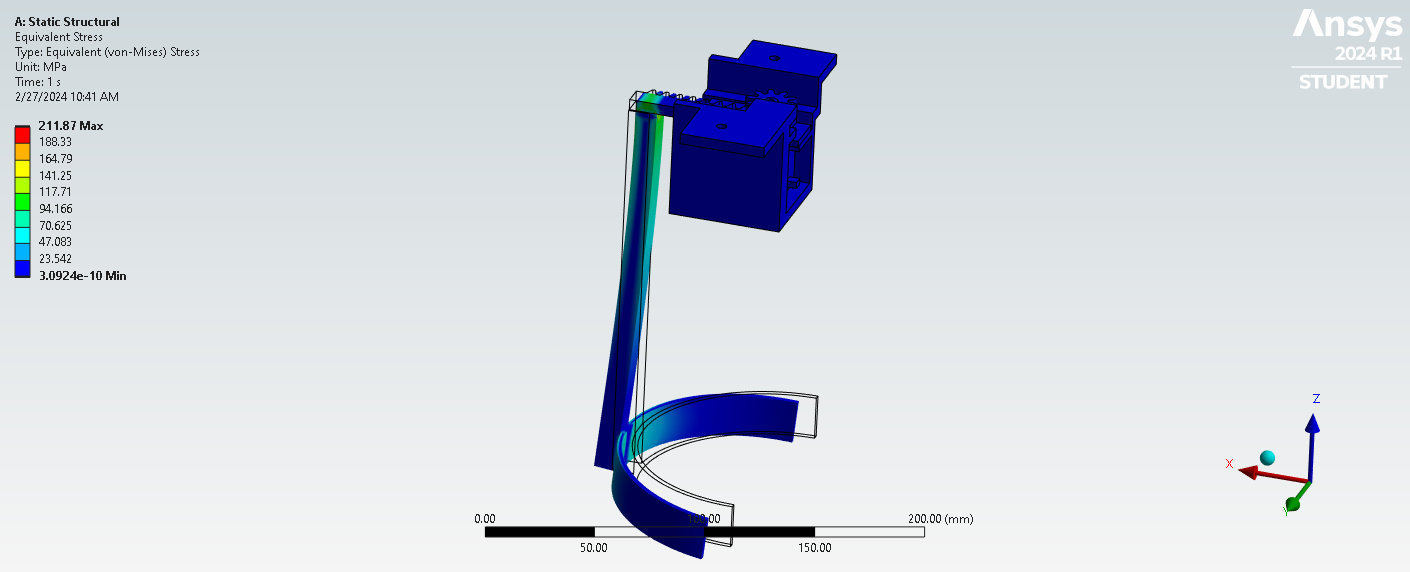}
\caption{Static structural simulation on rack representing equivalent Von-Mises stress}
\label{racksim2}
\end{figure}

Static structural analysis was conducted on the rack mechanism (see \hyperref[racksim1]{Fig. 22} and \hyperref[racksim2]{Fig. 23}) with bonded and friction-less contacts established between the bodies under investigation. Material assignments were made, utilizing ABS plastic for the 3D printed components and a combination of plastic and silicon for the motor. Quadratic and multi-zone meshing techniques were employed.

Fixed supports were applied within designated holes, while friction-less displacement supports were utilized for the rack and pinion mechanism. Additionally, a force was applied to the curved extension section, intended to displace the CanSat.

The resulting analysis yielded maximum equivalent Von-Mises stress of $211.87 MPa$ and maximum total deformation of $170.37 mm$. 

\section{Possible Failures in Mechanism}
\begin{itemize}
    \item The extension link could break due to insufficient force.
    \item The gears could be attached improperly during the assembly which can lead to them slipping and failing of mechanism.
    \item Irregular surface finish of the bulkheads can create hindrances during sliding of the payload.
    \item The batteries of the deployment system could fail leading to loss of power in the system.
\end{itemize}

\section{Concept of Operations (CONOPS)}
\begin{figure}[!htbp]
\label{conops}
\centering\includegraphics[width=1\linewidth]{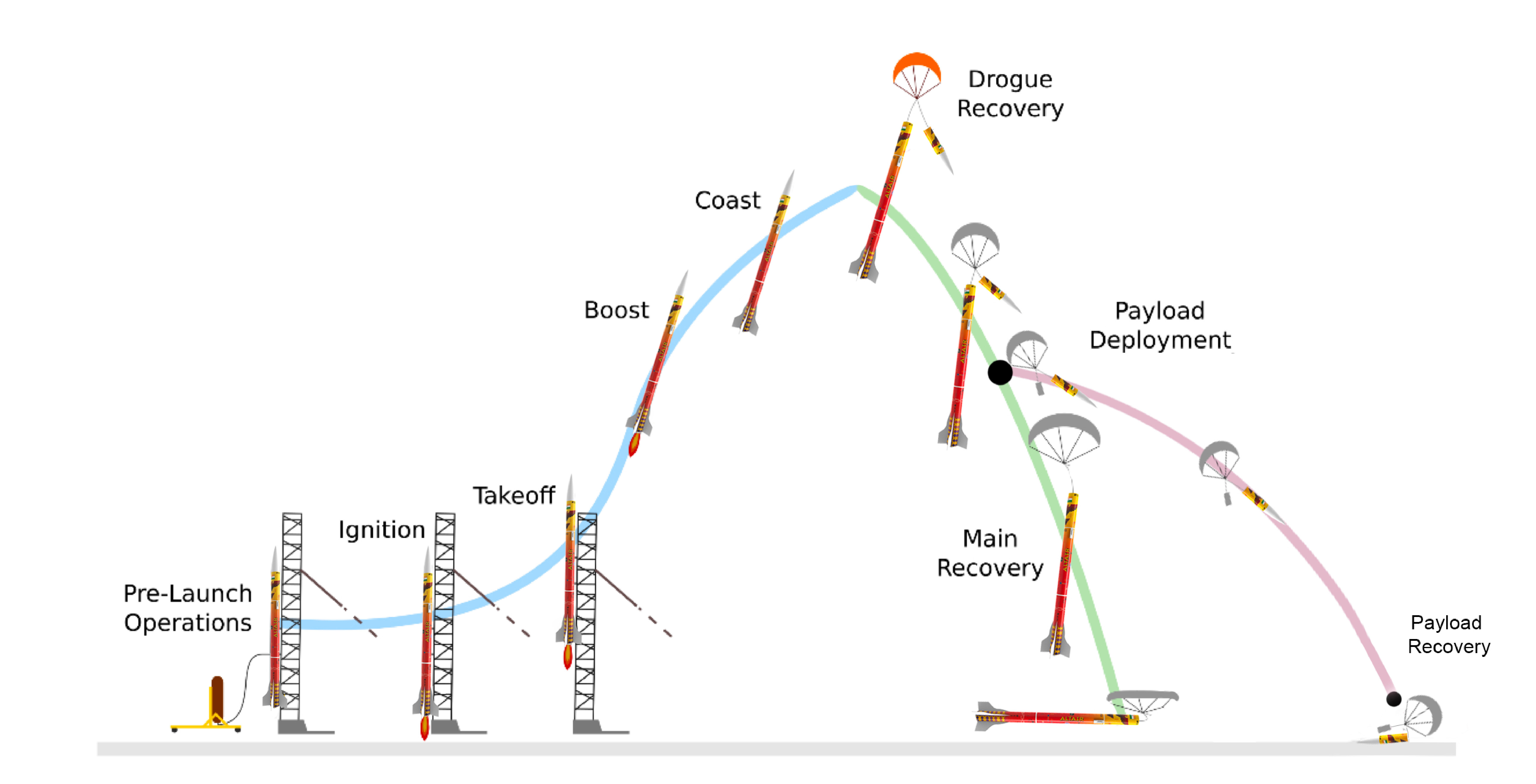}
\caption{CONOPS Stages}

\end{figure}

Following the ignition and lift-off, the rocket experiences a boost and coast phase reaching its apogee. At the apogee, the drouge parachute is deployed starting the recovery phase. During recovery, when the desired altitude for deployment of the payload is attained, a carefully controlled sequence of actions is initiated to facilitate the successful ejection of the CanSat payload. The process involved a series of well-defined steps, which are discussed in detail below:

1. \textit{Signal Transmission and Door Unlocking}: Upon reaching the apogee, the designated signal was transmitted, prompting the activation of the linear actuator responsible for unlocking the CanSat door. This critical step allowed for the subsequent release of the payload.

2. \textit{Carrier Mechanism Activation}: After the door's opening, a brief delay of 2 seconds was introduced to ensure the stability of the system. Subsequently, a pulse was sent to the carrier mechanism, employing the rack and pinion mechanism. This pulse initiated the continuous translation of the carrier mechanism three times, effectively pushing the CanSat payload outwards.

3. \textit{Radial Translation of CanSat Payload}: The rack arm applies the generated tangential force on the payload CanSat, resulting in the smooth radial translation of the CanSat payload away from the launch vehicle. This controlled radial movement is vital to ensure the safe and precise ejection of the payload.

4. \textit{Independent Descent of Payload}: Following its radial translation, the CanSat payload enters a state of independent free-fall. The parachute attached to the payload opens and the payloads descends steadily until it makes contact with the ground. This unassisted descent phase allows the payload to undergo its intended mission objectives without any further interference.

\begin{figure}[!htbp]
\label{flowchart}
\centering\includegraphics[width=0.7\linewidth]{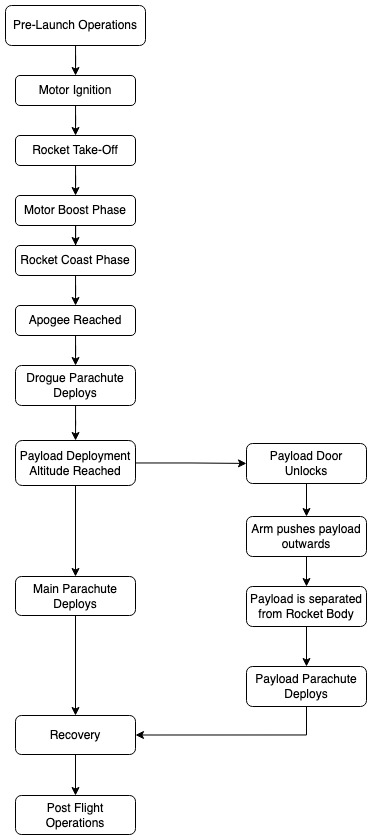}
\caption{Flow Chart Representing Radial Deployment Process}

\end{figure}

It is important to note that the successful ejection process and the subsequent payload descent are integral components of the overall mission's success. The precise execution of these actions guarantees the accurate deployment of the payload, ensuring the gathering of reliable data during its descent phase. A detailed visual description of the rocket flight, payload deployment and recovery recovery can be found in \hyperref[conops]{Fig. 24} and \hyperref[flowchart]{Fig. 25}. 

\section{Conclusions}
The research paper introduces a novel non-pyrotechnic payload deployment mechanism for sounding rockets, addressing a crucial challenge in the field. The mechanism is designed to be exceptionally suitable for sounding rockets, featuring a cylindrical carrier structure equipped with multiple independently operable deployment ports. The mechanism's radial deployment capability allows payloads to be released at different altitudes, thereby offering greater flexibility for scientific experiments. The paper presents the mechanism's design and conducts a comprehensive performance analysis. The findings demonstrate the viability and efficiency of this proposed mechanism for deploying multiple payloads within a single sounding rocket launch. The successful ejection process and the subsequent payload descent are integral components of the overall mission's success. The precise execution of these actions guarantees the accurate deployment of the payload, ensuring the gathering of reliable data during its descent phase. The mechanism represents a significant advancement in sounding rocket technology and holds promise for a wide array of applications in both scientific and commercial missions.

\section{Future Directions}

In the pursuit of refining and enhancing this innovative deployment mechanism, a series of avenues beckon for exploration. One such avenue entails the expansion of the mechanism's capabilities through thoughtful extensions. By extending the pushing rod, the mechanism could potentially accommodate stacked payloads, creating an avenue for multiple payloads to be released in tandem. This extended translatory motion holds promise for more sophisticated mission profiles and expanded scientific data gathering.

Furthermore, the door mechanism itself presents an opportunity for improvement. By integrating hydraulic components into the system, a controlled and deliberate lowering motion can be achieved, harmonizing with the unlocking of the latch. The unlocked door can also be used as a air-brake mechanism for soft landing of the rocket during descent phase. This heightened level of control, realized as the linear actuator is gracefully retracted, could potentially enhance overall mission dynamics and the precision of deployment.

In conclusion, the intricacies of the deployment mechanism have been successfully navigated, yielding a controlled and synchronized radial ejection of the CanSat payloads. The culmination of engineering ingenuity, empirical validation, and prospective enhancements underscores the significance of this mechanism in furthering the realm of payload deployment technologies.

\backmatter

\bmhead{Acknowledgements}

We sincerely thank thrustMIT, Manipal Institute of Technology, and the Manipal Academy of Higher Education for their invaluable support and resources that greatly facilitated the successful completion of this research. Additionally, we would be remiss to not highlight the contributions of \href{mailto:diyaparekh0603@gmail.com}{Diya Parekh}, and \href{mailto:hrishikesh.singh@learner.manipal.edu}{Hrishikesh Singh Yadav}. Their unwavering assistance and commitment to fostering a conducive research environment have been instrumental in shaping the outcomes of this study.

\section*{Statements \& Declarations}
\subsection*{Funding}
The authors declare that this research was conducted by authors as members of thrustMIT Rocketry Team at Manipal Institute of Technology, Manipal Academy of Higher Education (MAHE). All equipment utilized, expenses incurred during the research as well as the cost of publishing the manuscript is supported by Manipal Academy of Higher Education(MAHE) . 

\subsection*{Employment}
At the time of research, all the authors except the corresponding author are undergraduate students at Manipal Institute of Technology, Manipal Academy of Higher Education (MAHE). Corresponding author is employed as an Associate Professor (Senior Scale) in the Department of Aeronautical \& Automobile Engineering at Manipal Institute of Technology, Manipal Academy of Higher Education (MAHE).

\subsection*{Competing Interests}
The authors have no relevant financial or non-financial interests to disclose.

\subsection*{Author Contributions}
All authors contributed to the design, analysis and verification of the research produced. Design, development, testing and analysis of the mechanism were done by  Thakur Pranav G. Singh and Utkarsh Anand. The drafts of the manuscript was written by Tanvi Agrawal \& Utkarsh Anand and all authors commented on previous versions of the manuscript. Srinivas G. served as the project guide overlooking at design, verification and feasibility throughout mechanism development. All authors read and approved the final manuscript.

\subsection*{Consent for publication}
The authors declare that the manuscript being submitted is original and it not being considered for publication anywhere else. They consent for publication of the manuscript.

\subsection*{Data availability}
Not applicable

\subsection*{Materials availability}
Not applicable

\subsection*{Code availability}
Not applicable

\begin{appendices}

\section{Technical Documents}
\subsection{Mechanical Draft of Mechanism}
\begin{figure}[H]
\label{draft}
    \centering
    \includegraphics[width=0.8\textwidth]{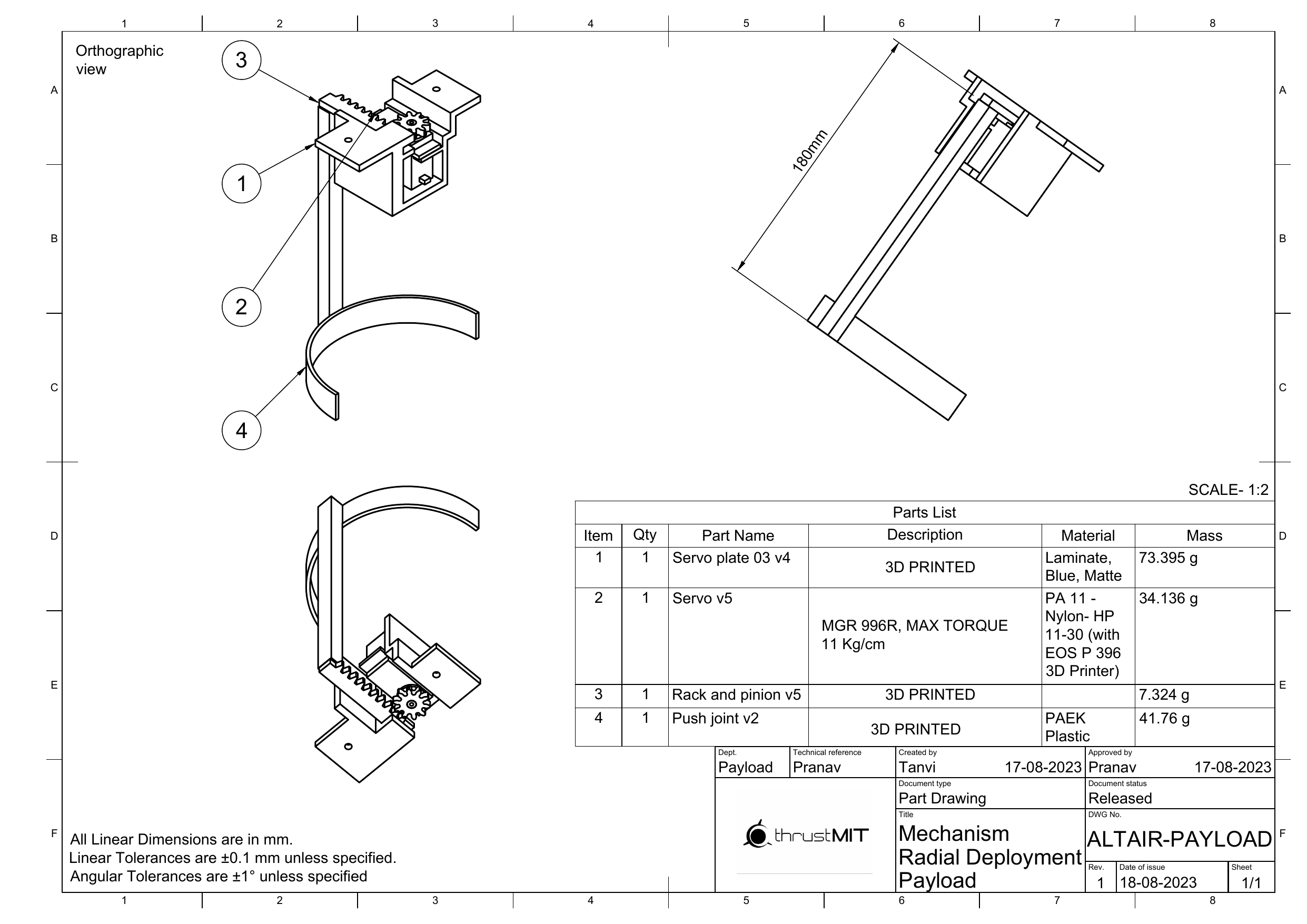}
    \caption{Draft of the Radial Deployment Mechanism}
    
\end{figure}
\subsection{Schematic of Electrical System}

\begin{figure}[H]
\label{pcb}
    \centering
    \includegraphics[width=0.8\textwidth]{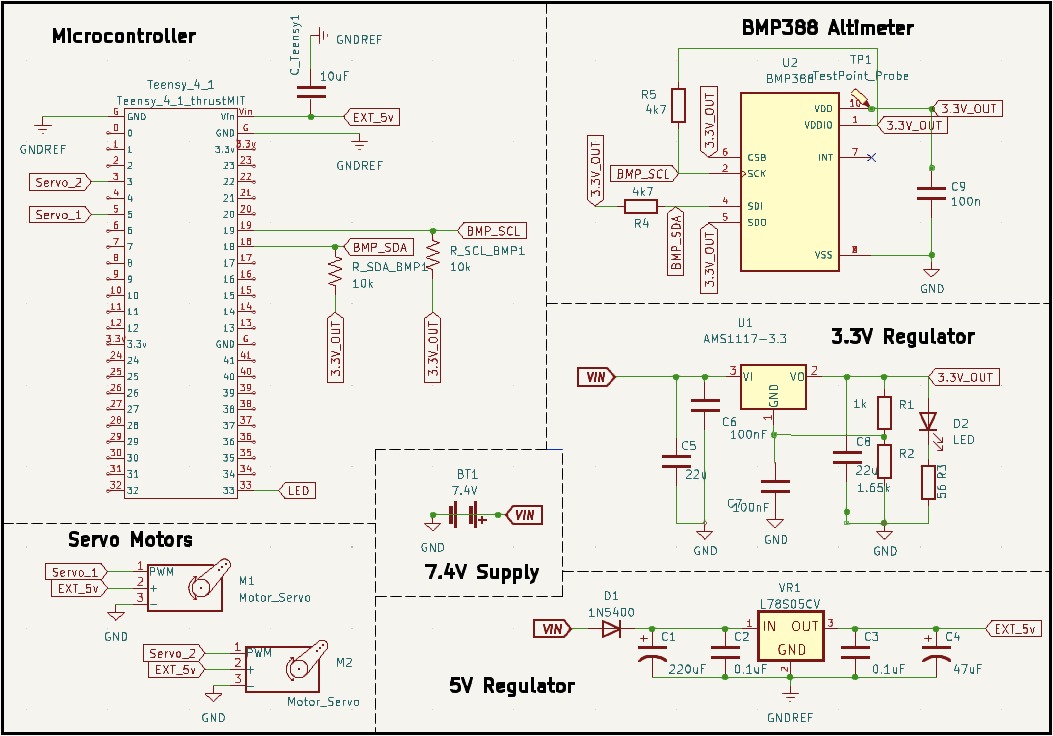}
    \caption{Schematic of Electrical System}
    
\end{figure}

\section{Abbreviation Used}
\begin{table}[!htbp]
\caption{Abbreviations Used}
\label{abb}
\centering
\begin{tabular}{|c|c|} 
\hline
3D & Three Dimensional\\
\hline
ABS & Acrylonitrile Butadiene Styrene\\
\hline
CAD & Computer Aided Design\\
\hline
CONOPS & Concept of Operations\\
\hline
ESA & European Space Agency\\
\hline
IC & Integrated Circuit\\
\hline
ISRO & Indian Space Research Organization\\
\hline
LiPo & Lithium Polymer\\
\hline
MEMS & Micro-Electromechanism System\\
\hline
NASA & National Aeronautics and Space Administration\\
\hline
RPM & Rotations Per Minute\\
\hline
RTOS & Real Time Opearting System \\
\hline

\end{tabular}

\end{table}




\end{appendices}


\bibliography{sn-article}

\end{document}